%% file: isaac.tex
\documentclass{scrartcl}
\pdfoutput=1
\usepackage{authblk}
\usepackage{url}
\usepackage{caption}
\usepackage{subcaption}
\usepackage{mathtools}
\usepackage[table]{xcolor}
\usepackage{amssymb}
\usepackage{adjustbox}
\usepackage{array}
\newcolumntype{M}[1]{>{\centering\arraybackslash}m{#1}}

\definecolor{mygreen}{rgb}{0,0.6,0}
\definecolor{mygray}{rgb}{0.5,0.5,0.5}
\definecolor{mymauve}{rgb}{0.58,0,0.82}
\definecolor{myblue}{HTML}{E5F2FF}
\usepackage{listings}
\newcommand\ack[1]{
	\vspace{1em}
	{\textit
		{#1}

	}
}

\begin{document}

\author[1,2]{Alexander~Matthes}
\author[1,2]{Axel~Huebl}
\author[1]{Ren\'e~Widera}
\author[2]{Sebastian~Grottel}
\author[2]{Stefan~Gumhold}
\author[1]{Michael~Bussmann}
\affil[1]{Helmholtz-Zentrum Dresden -- Rossendorf, Dresden, Germany}
\affil[2]{Technische Universit\"at Dresden, Dresden, Germany}

\title{In situ, steerable, hardware-independent and data-structure agnostic visualization with ISAAC}
\date{September 17, 2016}

\maketitle{}

\begin{abstract}%
The computation power of supercomputers grows faster than the bandwidth of their storage and network. Especially applications using hardware accelerators like Nvidia GPUs cannot save enough data to be analyzed in a later step. There is a high risk of loosing important scientific information. We introduce the in situ template library ISAAC which enables arbitrary applications like scientific simulations to live visualize their data without the need of deep copy operations or data transformation using the very same compute node and hardware accelerator the data is already residing on. Arbitrary meta data can be added to the renderings and user defined steering commands can be asynchronously sent back to the running application. Using an aggregating server, ISAAC streams the interactive visualization video and enables user to access their applications from everywhere.
\begingroup
\renewcommand\thefootnote{}\footnote{This work is submitted for publication to Supercomputing frontiers and innovations. The Supercomputing frontiers and innovations copyright notices apply.}%
\addtocounter{footnote}{-1}%
\endgroup


\end{abstract}

\input{content/introduction}
\input{content/related_work}
\input{content/system_design}
\input{content/evaluation}
\input{content/conclusion}

\ack{
ISAAC is open-source under the GNU Lesser General Public License (LGPL) Version 3+~\cite{isaacwebsite}.
This project has received funding from the European Unions Horizon 2020 research and innovation programme under grant agreement No 654220.
The talk about ISAAC at the ISC Workshop On In situ Visualization 2016 was founded by the Nvidia GPU Center of Excellence Dresden.
Furthermore we want to thank the PIConGPU contributors for their help including and evaluating ISAAC with their simulation.
}

\ack{This paper is distributed under the terms of the Creative Commons
Attribution-Non Commercial 3.0 License which permits non-commercial
use, reproduction and distribution of the work without further
permission provided the original work is properly cited.}%

\bibliography{citations}

\end{document}

%% file: content/introduction.tex
\section{Introduction} \label{sec:introduction}
Supercomputers are getting faster every year. 2016 the National Supercomputing Center in Wuxi released its Sunway TaihuLight with over 93 PFLOP/s peak performance in the LINPACK Benchmark~\cite{MSD+16}. The system provides over $1.3$ petabytes main memory enabling scientists having a glimpse at the upcoming exascale area.

Unfortunately neither the mass memory, nor the network interconnect, not even the system's internal interconnect (e.g. PCIe) are fast enough to be able to transfer or save all data created in high performance computing (HPC) applications, e.g. scientific simulations, running on such systems without unacceptable latencies. Consequently only a small subset of the processed data can be analyzed afterwards with the classical post processing approach which means loosing maybe important scientific data.

One solution is to analyze the data on the same system they are produced on while they are produced, so called in situ processing. The benefits are the reduction of the system's bandwidth bottlenecks, especially the network disk bandwidth, and the instantaneous feedback from the HPC application. It becomes apparent that analyzing the data of the application without going in situ will not be possible in the age of exascale computing anymore~\cite{ma2007situ}. However, in situ processing creates new challenges: A connection between application and the in situ process must be enabled, maybe even the application changed. Furthermore, using live in situ analysis, the user cannot just queue an application anymore and forget about it until it is done, analyzing it sometime afterwards. The scientist needs to be connected to the application while its running to steer the analysis process. Data not been analyzed at this point are gone. Last but not least HPC clusters are not suited for a direct connection of observing and steering users from the outside. You cannot decide or know beforehand, on which compute nodes your job will be executed. Additionally, most of the time computation nodes are not accessed from outside the cluster. For most systems a direct connection is not possible at all.

In this paper we present ISAAC, an open-source, in situ visualizing and steering library which addresses the stated isssues. ISAAC does not only enable arbitrary applications to create volume renderings of their scientific data, but also establishes an easy to use way of passing and receiving messages from and to the application. Using C++ metaprogramming techniques no deep copy of already existing data is needed; The visualizing algorithm can work on the original data of the application itself, but is still independent of domain-specific layouts, alignments or shifts of the underlying data.

Right now six different architectures are present in the world wide top 10 of the supercomputers, including the well-known X86, Sparc, or PowerPC CPU architectures, but also hardware accelerators like Nvidia Tesla or Intel Xeon Phi. The already mentioned TaihuLight introduces a new architecture SW26010. To be able to interact with all these different architectures without loosing performance portability, ISAAC uses Alpaka, an open-source library for the zero overhead abstraction of highly parallel CPUs and hardware accelerators~\cite{zenker2016alpaka}. Alpaka abstracts from different parallel programming models like CUDA, OpenMP or std::thread to a single programming model, similar in structure to CUDA, enabling host CPUs to be abstracted as hardware accelerators. In the future also support for OpenACC, Intel Thread Building Blocks or AMD HIP is envisaged. With this approach ISAAC can concentrate on a single programming model while still being able to run on every hardware of the top 10 systems. In this work we will concentrate on one of these systems, Piz Daint, to test the performance of ISAAC in a real-world situation.

For this we have used ISAAC to visualize the open-source plasma simulation PIConGPU running on GPUs~\cite{bussmann2013radiative,burau2010picongpu}. For the evaluation 4096 Nvidia Kepler GPUs of the Piz Daint cluster are used. We show that ISAAC is capable of visualizing a highly parallel petascale application and scales for the upcoming exascale era.

%% file: content/related_work.tex
\section{Related Work}
\label{sec:related}

In situ processing and especially visualizing is a well investigated topic, not only presenting some theoretical thoughts, but also with a lot of ready to use libraries and tool sets.

Besides visualizing the application data, other data compressions are possible. Lakshminarasimhan et al. showed that a lossy compression of $85 \%$ of scientific data is possible with a guaranteed correlation of $>0.99$ with the original data\cite{lakshminarasimhan2011compressing}. The libraries zfp and zplib have similar approaches for lossy and lossless compressing of floating point arrays and 2D and 3D fields to decrease the amount of data to save or transfer~\cite{lindstrom2006fast,lindstrom2014fixed}. Another method is to filter the data before saving or transferring it for a classical post processing with only using a representative sample of the whole data~\cite{woodring2011situ}.

Literature furthermore differs between the distance of computation and visualization~\cite{bennett2012combining}. Tightly coupled in situ processing uses the very same hardware whereby loosely coupled processing uses dedicated but with a fast interconnect connected visualizing nodes. To differentiate these two, the last one is also called in-transit processing~\cite{bennett2012combining}. The tight coupled approach has the benefit of reusing the simulation data without the need of copying~\cite{tu2006mesh,fernandes2015situ,yu2009study}. Internal application data and functions can be used for a faster visualization as shown from Yu et al.~\cite{yu2006remote}. Apart from that, application and visualization then compete for the same resources. Load balancing may be needed for the most efficient usage of the hardware~\cite{hagan2011multi}. Hybrid systems first filter and compress data before visualizing them loosely coupled~\cite{rivi2012situ}.

Especially hardware accelerators often have only limited memory compared to the host system which is already fully used by the HPC application. The analysis of application data often requires creation of consecutive transformations of the raw data, often stored in a new temporary buffer. Beside wasting memory not usable from the HPC application anymore, this approach also does not allow CPUs to cache data before the next analysis step. Most of the cycles are used writing or reading data to and from memory. Moreland et al. describe a much faster approach defining a chain of so called \emph{worklets} describing the transformations which are executed on every date from the raw data to the final transformation without storing it in temporary buffers~\cite{moreland2011dax}.

As intrinsic challenge of in situ, users have normally to be connected to the HPC application to be able to steer it. Kageyama et al. use a different approach called Light-Field-Rendering, in which the application is visualized from dozens or even thousands of different directions and the videos saved automatically. Using modern compression algorithms like H264 the produced videos are much smaller than the raw application data and much easier to be saved to hard disk. Later the user can analyze the visualization choosing and interpolating the correct video based on the viewing angle~\cite{kageyama2014approach}.

Ahrens et al. implemented a similar solution called ParaView Cinema. Based on options set right before the run, like camera position and angles, iso surface thresholds, cutting planes, and similar visualization parameters, a database with millions of images is created live in situ. Although this database is multiple terabytes big, it is still smaller than saving the raw scientific data directly. The post processing application then can query for specific images similar to the approach of Kageyama et al., but as no videos but an image database is created, the iso surface images can consist, e.g., of normals and the hit values instead of colors, enabling the user to adjust coloring and lighting interactively~\cite{ahrens2014image}.

ParaView~\cite{cedilnik2006remote} and VisIt~\cite{childs2013visit} are two of the most used tools in use for scientific visualization. As they are open-source, other groups were able to extend them to enable in situ rendering~\cite{rivi2012situ}. The library Libsim implements the tightly coupled approach for VisIt~\cite{kuhlen2011parallel}, whereas two in-transit approaches exist for ParaView: As most scientific applications are able to write to the de facto standard for scientific data, HDF5, the library ICARUS implements a HDF5 driver which sends the data to a ParaView visualization server instead of being written to a hard disk~\cite{biddiscombe2011parallel}. The built-in library ParaView Coprocessing of ParaView uses a similar approach, but defines a C-like interface like Libsim which needs to be used from the HPC application\cite{fabian2011paraview}.

Beside these general solutions also some attempts have been made, directly adding in situ visualization to already existing application codes~\cite{ma2009situ,ning2010design}. The benefit is in the possibility to directly reuse the existing data structures. Especially applications running on hardware accelerators don't need to deep copy the data to the host interface of the generic in situ libraries\cite{stone2013gpu}. Although being very fast these solution have the disadvantage of being very hard to port to another application. In fact one would just rewrite the visualization code.

%% file: content/system_design.tex
\section{System design}
\label{sec:design}

The goal of ISAAC is to combine the benefits of a generic library solution with the speed of custom-built in situ visualizations. For this ISAAC is implemented as a header only library meaning that the included C++-header does not only contain the interface but the whole library code and thus does not need to be linked against in the end. With this approach we can use the same types defined and used by the HPC application and the underlying hardware whereby no data conversion is needed at all.

In its current version ISAAC provides the following visualization features
\begin{itemize}
\item volume rendering, 
\item iso surface rendering,
\item clipping planes,
\item freely definable transfer functions,
\item data interpolation,
\item zero data copy capability with the ability to directly operate on the simulation data,
\item freely configurable data transformation using functors and
\item a server-client architecture allowing for simulation steering and remote visualization.
\end{itemize}

\subsection{Design concept}

The idea behind ISAAC is to work on the simulation data in-memory. This means that the structure of simulation data is transformed in a way to be suitable for visualization without the need for copying the data. This is important to reach optimum performance for current multi- and many-core compute hardware. On such systems, the fastest memory with the highest bandwidth and the lowest latency usually is small compared to the overall memory available in a node. Deep copies between levels in the memory hierarchy, for example from the accelerator memory to the main memory of a node, usually come with a degradation in performance.

With it's zero copy philosophy ISAAC operates directly on the simulation data at the highest bandwidth and lowest latency available while keeping memory consumption at a minimum. It makes extensive use of C++ template metaprogramming techniques in order to transform data by reindexing during compile time and abstracting parallel computations using the library Alpaka, see section \ref{sec:introduction}.

With this, ISAAC can define its rendering algorithm completely independent from the using application. Instead of defining a C library interface like Libsim or ParaView Coprocessing, ISAAC defines a type independent interface which the application needs to feed with accessors to the real data which handle the alignment and pitch, but also can perform a reordering of the domain dimensions. The accessor itself can be removed at compile time (zero overhead abstraction).

Instead of deciding on which data transformations to apply at runtime using a large loop which includes all data sources and visualization features and selecting them based on user input ISAAC unrolls this loop, compiling all possible combinations of features during the generation of the final executable. This is important to allow for hardware-dependent optimization at compile time and for keeping branch divergence in the code to a minimum. Uisng the functor chain concept introduced later this enables the user to choose arbitrary visualization features at runtime while providing for optimum code performance.

To be able to interact with arbitrary application code, ISAAC abstracts the HPC application describing a cuboid-shaped 3D volume, in which each compute node has its own disjoint, also cuboid-shaped local part of the global volume. Applications often use copies of other local volumes of other nodes called \emph{ghost regions} or \emph{guards}. These regions are exploited from ISAAC for interpolation over node borders, but are not part of the disjoint volume and thus not directly accessed in the ray casting. The data calculated in the application are abstracted as \emph{sources}. A source assigns a scalar value or a vector for every regularly spaced position in the global volume defining a field this way. Thereby every compute node only needs to know the assignment for its local sub volume.

\begin{figure}[tb]
	\centerline{
		\resizebox{1.0\textwidth}{!}{\includegraphics{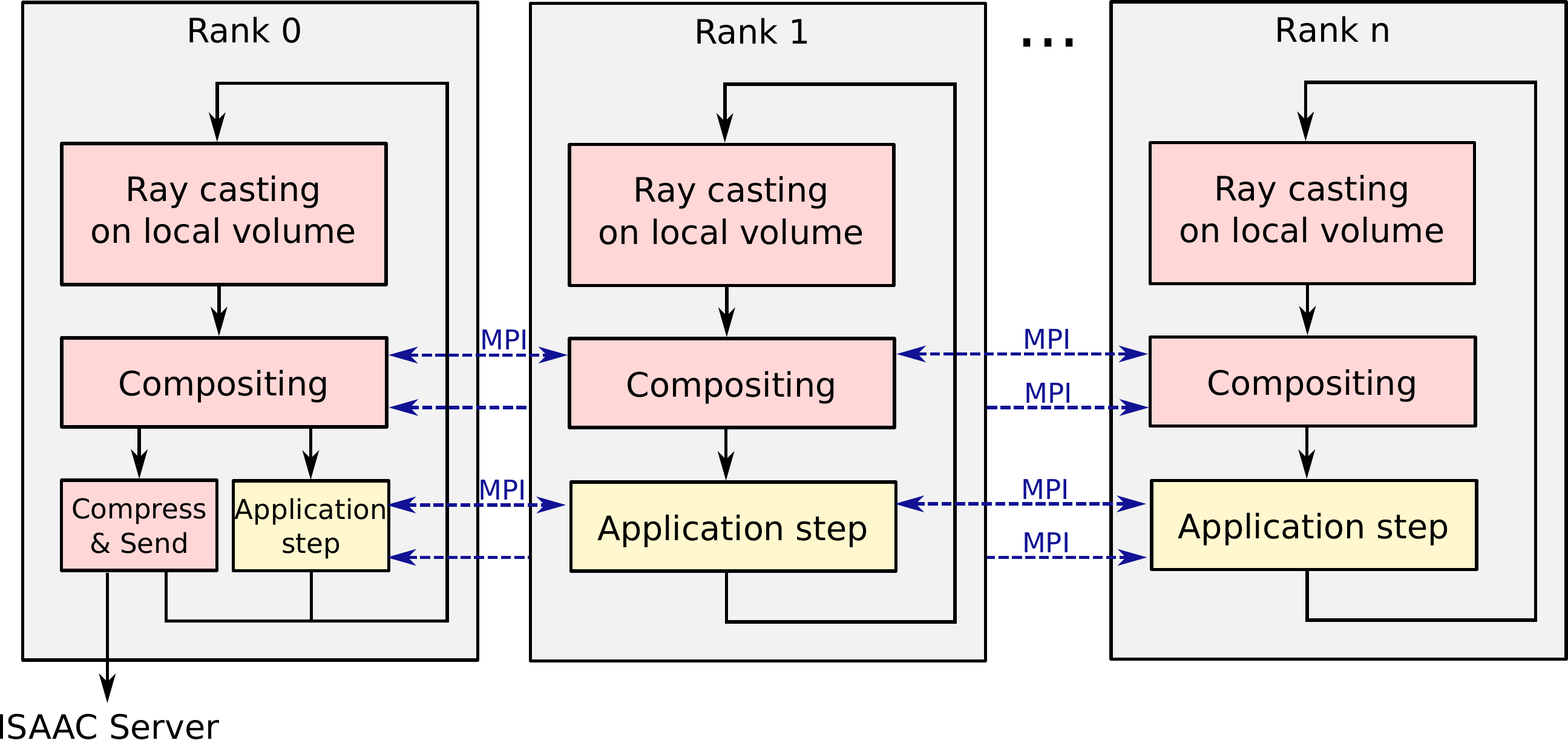}}
	}
	\vspace{3mm}
	\caption{Whole concept of ISAAC. First a local ray casting is done. The resulting images are composited using the interconnect. The whole image is only on the first node in the end. Afterwards the application gets back the focus on every node, but on the first node the image is also compressed and sent to the ISAAC server in parallel.}
	\label{fig:whole_concept}
\end{figure}

On the local volume ISAAC performs a ray casting over all sources on every node. The local images of all nodes are then composited. Afterwards the application gets back the focus and ISAAC returns except for the first node on which the image is compressed and sent to a central streaming server in the background as seen in figure~\ref{fig:whole_concept}. The details will be described in the following sub sections.

\subsection{Class based interface}

\begin{figure}[tb]
\begin{lstlisting}[caption={Definition of a source in ISAAC}, label={lst:source_definition}, escapeinside={|}{|}]
struct TSource
{
	//Feature dimension of the source. 1 => scalar field
	static const size_t f_dim = 1;
	//Option, whether the algorithm may read data outside the devices local volume
	static const bool has_guard = true;
	//Non persistent data need to be copied before rendering
	static const bool persistent = true;

	//Name of the source
	static std::string getName()
	{
		return std::string("Test Source");
	}

	//Function to be called for every source right before the rendering starts
	void update(bool enabled, void* pointer) {}

	//Accessor for the field data of the source
	float_vec<f_dim> operator[] (const int_vec<3>& nIndex) const
	{
		float_vec<f_dim> result = { 42 };
		return result;
	}
};
\end{lstlisting}
\end{figure}

Listing \ref{lst:source_definition} shows a local description of a source. An application defines such a class for every source. It consists of three static, constant attributes and three member functions. The attribute \texttt{f\_dim} defines the dimension of the vector field called \emph{feature dimension}. ISAAC supports vectors with up to 4 dimensions. A vector field with more than four dimensions could be implemented as two or more sources with smaller dimensions sorted in semantic groups. The data accessor \texttt{operater[]} is used to assign a value for an integer index position \texttt{nIndex} in the local volume of the compute node. In this example the accessor does not have to read from memory at all. Even an analytic description of a source is possible. This example source, e.g., defines a homogeneous scalar field which is $42$ everywhere. 

ISAAC works on a 3D location domain, but it is still possible to define sources for 2D applications with just ignoring the z component of the position. If possible applicationwise, past timesteps could be mapped to the z component using the time as third dimension.

Although the ray casting algorithm does only work on the local volume, for interpolation between the integer positions it might be handy to access the most outer region of a neighbor volume residing on another device or compute node. ISAAC itself does not communicate these guards or ghost regions between the nodes, but can use it if the underlying application transparently provides those. To tell ISAAC that such a region exists outside the local volume, the flag \texttt{has\_guard} deactivates a border check if interpolation is activated.

Right before ISAAC starts the ray casting for each source the feedback function \texttt{update} is called which can also be used to forward frame depended informations (like the time step) to the sources. This function shall be used to prepare the accessors, so that they can deliver correct fields when the rendering begins. ISAAC is primary meant to visualize already existing fields which are called persistent. But an application can also create secondary fields just for the visualization at this point. As some applications have only one temporary buffer for such fields, but maybe more than one source using it, it is also possible to tell ISAAC with setting the flag \texttt{persistent} to false to make a copy of this field using the data accessor, so that the update functions of the following sources are safely able to reuse the temporary buffer.

\begin{figure}
	\centerline{
		\resizebox{1.0\textwidth}{!}{\includegraphics{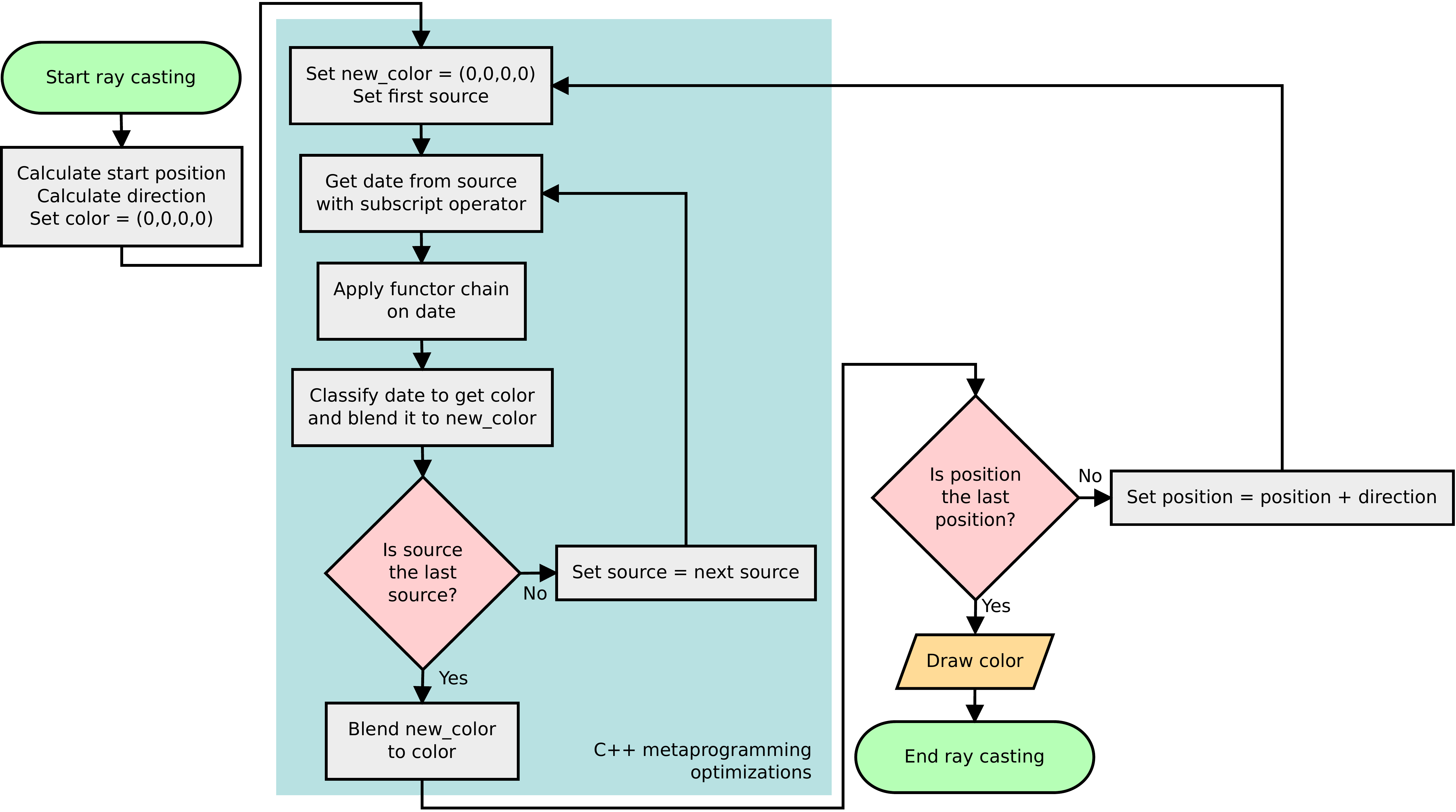}}
	}
	\caption{Simplified flow chart of the ray casting of ISAAC. The inner loop over all sources is evaluated at compile time.}
	\label{fig:pap}
\end{figure}

For every scan point of the ray casting, the volume rendering algorithm iterates over all sources, checks for the flags \texttt{has\_guard} and \texttt{persistent}, uses the \texttt{[]operator} (subscript operator) to get a date from the source, applies a functor chain (see next sub section), classifies it to a color and opacity, and finally uses these to calculate a global color and opacity for the scan point to be used in the classical front to back ray cast algorithm. Figure~\ref{fig:pap} shows all renderings loops and the metaprogramming optimizations as flow chart.

Iterating over all sources for every scan point for every pixel of the resulting image and the checks for the flags in every round would add an enormous looping and comparison overhead. To avoid this, all sources need to be defined at compile time and the iteration is done via C++ metaprogramming. The whole loop over all sources is evaluated and unrolled at compile time and does not produce overhead at runtime. Furthermore, the \texttt{static const} flags can be evaluated at compile time and the compiler can optimize the data access having detailed informations about the data accessors and their arithmetical connection. Also if different sources use the same or very close data in the background, the compiler can optimize the data access such that it is read only once although it may be abstracted differently sourcewise.

However, looping over all sources at compile time has the big disadvantage that every source is touched, even if it may not be of current interest at all. Even a classification assigning the opacity zero to every position in the local domain would still read the source data although it is discarded in the end. To avoid this, $2^n$ different render functions are created at compile time for every combination of activated sources, where $n$ is the total number of sources. Although the rendering functions are generated at compile time, ISAAC chooses the right function at runtime based on interactive user settings. A deactivated source is not touched at all, but we still have the benefit of looping over all activated sources at compile time.

Although this approach is quite fast, the number of needed rendering functions grows exponentially especially for big $n$. However, in the field $n$ should not be bigger than $10$ for most applications resulting in a manageable amount of $1024$ rendering functions. If an application defines more sources, only preselecting the most interesting ones at compile time is possible for now.

In the future ISAAC could be improved in such a way that $n$ sources may be defined, but only $m$ activated at the same time which can be calculated with the binomial coefficient ${n \choose m} = {{n!} \over {m! \cdot (n - m)!}}$ instead of with $2^n$. For $n=20$ and only three sources of interest at the same time ($m=3$) this would resolve to ${20 \choose 3} = 1140$ instead of $2^{20}=1048576$ render functions.

To finally use ISAAC, an application has to define the mentioned classes for every source, enqueue them in a C++ metaprogramming sequence and pass this list with some additional options as template parameters to the main ISAAC visualization class providing an optimized solution for the specific application without ISAAC needing to know anything about the application at all.

An instance of this class can then be used for creating visualizations of the HPC application. Every time a frame shall be sent to the user, just the render method of the object needs to be called. ISAAC will get the focus from the application, create the rendering and return afterwards. With this approach the developer has total control over ISAAC and no unpredictable behavior is happening in the background interfering with the application, see also figure~\ref{fig:whole_concept}.

\subsection{Functor chains}

ISAAC works on the original application data to avoid any deep copy in the main memory. However, not every application data is suited for a direct visualization. The transfer function from the date in the volume to a color and opacity needs a value range to work on. Of course this range could be set as \texttt{static const} option like for the guard and the persistence, but sometimes even the application users cannot estimate what the maximum range may be, as it can depend on run time parameters or non-linear processes.

Furthermore, the classification works on scalar values, but often applications have vector fields, too, and it is not obvious before running the application and actually seeing the data which dimension or transformation from a vector value to a scalar value fits best.

As solution to this problem, ISAAC uses an approach similar to the worklets of Moreland et al.~\cite{moreland2011dax}. We define a small set of very basic functions, called \emph{functors}, like a multiplication or addition with a constant vector or scalar value, calculating the length, or summarizing all vector elements to one scalar value. These functors can be concatenated with the pipe sign \texttt{|} to more complex functions called \emph{functor chains} successively applying functors to a date. At the moment functor chains are limited to work on one source only. Table \ref{tab:functors} shows the predefined functors of ISAAC. Notice that the functor may change the domain of the input vector or scalar value to a vector of a different dimension or a scalar value.

\begin{table}[tb]
	\centering
	\begin{footnotesize}
	\begin{tabular}{|c|M{0.13\textwidth}|M{0.13\textwidth}|M{0.13\textwidth}|M{0.13\textwidth}|M{0.13\textwidth}|}
	\hline Predefined functors & \texttt{add(v)} & \texttt{mul(v)} & \texttt{length} & \texttt{sum} & \texttt{pow(v)} \\ 
	\hline Functionality & Adds the vector \texttt{v} & Multiplies the vector \texttt{v} & Calculates the length & Summarizes all vector components & Exponentiates componentwise with \texttt{v} \\ 
	\hline Domain change & $\mathbb{R}^n \rightarrow \mathbb{R}^n$ & $\mathbb{R}^n \rightarrow \mathbb{R}^n$ & $\mathbb{R}^n \rightarrow \mathbb{R}$ & $\mathbb{R}^n \rightarrow \mathbb{R}$ & $\mathbb{R}^n \rightarrow \mathbb{R}^n$ \\ 
	\hline 
	\end{tabular} 
	\end{footnotesize}
	\vspace{3mm}
	\caption{Five predefined functors in ISAAC for addition, multiplication, length, summarization and exponentiation. Functors may change the domain of the vector or scalar value and may have constant arguments.}
	\label{tab:functors}
\end{table}

Such a functor chain could be e.g. \texttt{mul(2,3,4) | add(1) | length}. This exemplary function takes the raw date from the source, multiplies with vector $\Big( \begin{smallmatrix} 2 \\ 3 \\ 4 \end{smallmatrix} \Big)$, adds the vector $\Big( \begin{smallmatrix} 1 \\ 1 \\ 1 \end{smallmatrix} \Big)$ and calculates the length of the resulting vector in the end returning a scalar value. If no reduction to a scalar value is done explicitly, the first component of the resulting vector is used by default for the following classification in the ray casting.

If only selected vector components are of peculiar interest, the functor \texttt{sum} can be used in this way: \texttt{mul(0,1,0) | sum}. In that case the first and third component get extinguished and afterwards all values are summarized to a scalar value representing the second component of the original vector.

With these functors most of the needed transformations for sources can be described. If a functor is missing, it can easily be added in the code at compile time. Internally the functors are just classes which need to implement four versions of a method for all four possible feature dimensions, e.g. using templates. A needed dimension reduction (or enhancement) can be expressed via the return type of the methods. These classes are enlisted in a C++ metaprogramming sequence and parsed at compile time. Without changing ISAAC itself, an application developer can add a functor especially useful for a specific application domain.

The functor chains can be defined at run time and they are executed right after reading the data from the source. The result of the functor chain execution is then used in the classification. However, checking for every functor of any source in the ray casting loop would be again slow because of the lookup overhead -- and useless as the functor chains do not change while rendering the image. Unfortunately it is not possible to create an own render function for every functor chain combination as every source can have a different functor chain. If we limit the maximum count of functors usable per source to $c$, this would create $f^c$ different combinations per source and ${f^c}^n$ combinations of combinations for all $n$ sources. Even without involving the already mentioned $2^n$ render functions, if we choose the quite small values $f=4$, $c=3$, and $n=3$ we would already have $262144$ render functions. With $n=5$ it would be over one billion.

Instead we compile one function for every possibly combination of functor chains for the total number of functors $f$ and some user defined maximum functor chain length $c$. At run time, if the functor chain is changed, the corresponding function pointer of the precompiled function is chosen and set up for the source. This has three benefits: Only $4 \cdot f^c$ functions need to be generated. The factor $4$ comes from the four different feature dimensions possible in ISAAC. For the mentioned examples $n=3$ and $n=5$ this solves to $256$ or $4096$ needed versions, whereby $n>c$ is questionable anyway as this would mean that one functor is used twice which should not be needed. But even with this unrealistic number this can be handled by the compiler in finite time.

With using function pointers the compiler cannot do cross-function optimization at this point anymore. We tried to exploit as much compile time optimization as possible but had to draw a line at one point which is reached here with creating thousands of already fused functor chains. In the field a further integration of the functor chains into the rendering function would have only a small or even no impact on the rendering time, but increase the compile time radically.

A CUDA specific optimization makes sense for the functor chains nevertheless. Alpaka abstracts the hardware accelerators in such a way that all available parallelization layers and performance gainers are part of the interface, but ignored for hardware not supporting them. ISAAC uses this to store the functor chain function pointers and the parameters of the functors in constant memory if the hardware supports this. Especially on Nvidia GPUs this improves the performance of the rendering noticeably.

\subsection{Parallel Rendering}

\begin{figure}
	\centerline{
		\resizebox{0.8\textwidth}{!}{\includegraphics{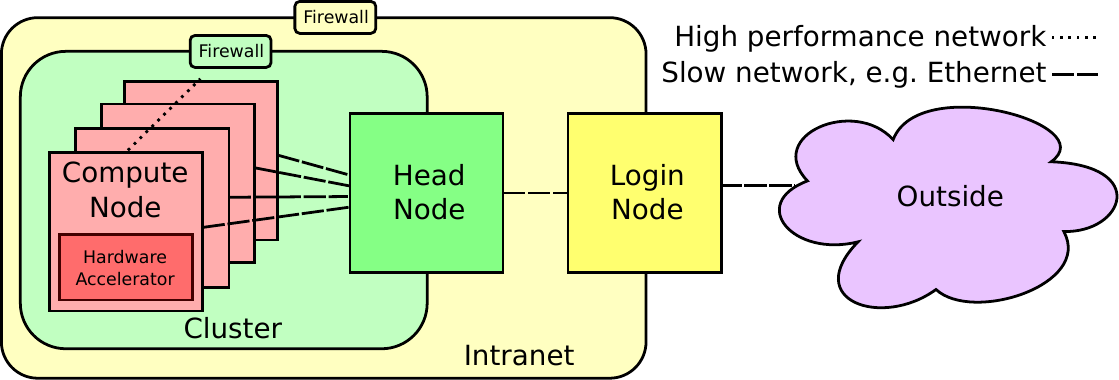}}
	}
	\caption{Simplified design of a typical cluster with multiple firewalls. The only connection between the compute nodes and the outside (e.g. the Internet) is over the cluster's head and login nodes. Only between computation nodes a high performance network can be assumed. Green shows the cluster, yellow the intranet of the site and purple the outside.}
	\label{fig:cluster}
\end{figure}

A typical cluster design can be seen in figure \ref{fig:cluster}. The heart of a cluster consists of a high number of compute nodes doing the main computation tasks. Among themselves they shall be connected over a high performance network like InfiniBand. Sometimes they are also connected to the head node via such a network, but often only a slower connection, like Ethernet, is available. The task of cluster head nodes is to manage the jobs running on the cluster and sometimes for compiling the cluster applications before distributing them to the compute nodes. Often the cluster itself is just a small part of a bigger IT infrastructure of a company or research site. To access this network from the outside, login servers are used. However, the connection inside the global site network is only using Ethernet most of the times which is fast enough for most activities, but cannot handle the raw data produced in a modern cluster. ISAAC is designed to work well in those environments.

After the local image of every local domain is rendered, ISAAC composites one big image out of it using the compositing library IceT~\cite{SNL11}. IceT composites the renderings over MPI using balancing techniques like binary swap~\cite{ma1994parallel} or 2--3 swap~\cite{yu2008massively}. These algorithms ensure that the network load and the computing load are well distributed over the whole cluster.

IceT supports collecting the composited image on a node not involved in a rendering of a sub domain at all. As the merged image needs to leave the cluster sooner or later, it had been considered collecting the final image on the head node. However, two problems occurred with this idea: First of all IceT works with MPI by default, whereas the head node is not part of the clusters MPI world and often of a different architecture. This could be solved with extending the open-source library IceT, but would mix MPI and Ethernet socket calls in IceT. Furthermore the user does allocate time on the compute nodes -- not the head node. It is not meant for being an important part of the HPC application itself.

\begin{figure}
	\centerline{
		\resizebox{1.0\textwidth}{!}{\includegraphics{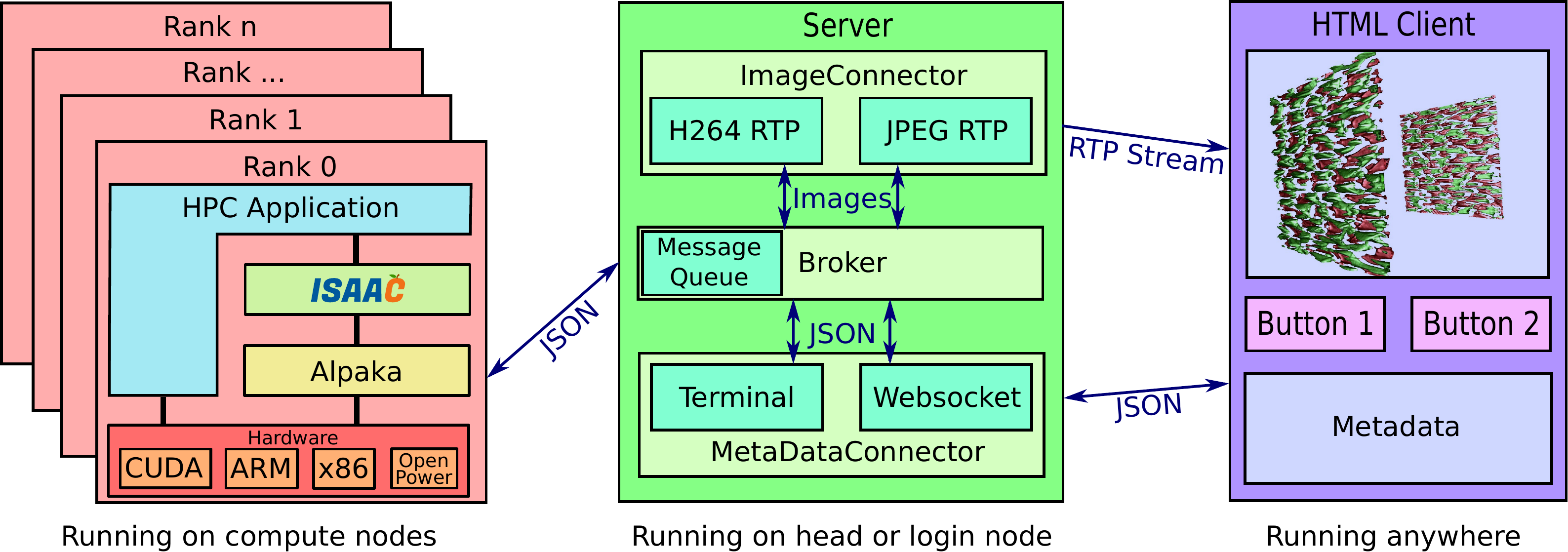}}
	}
	\caption{ISAAC consists of a library running integrated with the application on the compute nodes, a server running on one the head or login nodes, and an arbitrary amount of clients running anywhere. With this a connection between the compute nodes and the outside of the cluster can be established. Furthermore, the server can compress the data for slower networks.}
	\label{fig:concept}
\end{figure}

Instead the image is collected in the very first compute node (MPI rank 0). However, it is not good practice to let computers outside the cluster connect to the compute nodes ensured by internal firewalls. Considering this, ISAAC introduces a central server as seen in figure \ref{fig:concept}. Besides forwarding and managing connections of ISAAC sessions on the cluster and to an arbitrary amount of clients, it also has the possibility to compress the images and, e.g., to create streams. At the moment the ISAAC server supports creating RTP streams with H264 and JPEG encoded videos and RTMP for streaming services like Twitch or Youtube, but defines a simple interface to add other or future streaming services, too.

\subsection{Steering arbitrary HPC applications using open standards}

The connection between the server and the ISAAC library is using TCP/IP. To be able to exchange library or server or to extend the behavior of both, every message is formated in the easy to use and well documented \emph{JavaScript Object Notation} (JSON)~\cite{json16}.

The server is designed in such a way that not only the streaming service, but also the client connected to the server can be exchanged. We provide a ready to use HTML5 reference client which uses Websockets to establish a TCP/IP connection to the server. Again, JSON is used for an easy extensibility. As JSON directly defines a JavaScript object, using the received objects in HTML5 and sending feedback back to applications is straightforward. The client can easily be adapted to the specific needs of an application.

All JSON commands sent and understood by the ISAAC library and the server are documented~\cite{isaacjsondoc}. Additionally a \texttt{metadata} object is defined which can be used from the application as root object for arbitrary informations which shall be added to an image after it is created and before it is sent to the clients. ISAAC does not care for the content of this object, but ensures that the resulting JSON object is still valid with removing objects with the same name as every node can add its own meta data which are merged by ISAAC before being sent to the server. Nevertheless, it is possible to gather informations of all instances of the application binary running on the cluster using JSON arrays which are just concatenated by ISAAC.

The same way a client can send arbitrary informations or requests to a running application. The application gets ready-to-use JSON objects with the steering data of all connected clients. With this the application can be paused at a state of interest, to investigate the data, modify the functor chains or even change the application parameters themselves, e.g. decreasing the step width in a numerical integration. But also unsuccessful application runs can be identified early and corrected or restarted with better settings. Especially as jobs often need a long waiting time on a cluster before they are started, an application restart inside the job can decrease waiting times and improve productivity of scientists.

\subsection{Exploiting heterogeneous systems}

As the root rank and the server are often not connected with a high performance network, sending a raw bitmap takes more time than compressing the image and sending the smaller image. Because of this, a JPEG compression (with adjustable quality) is done on the root rank. The JPEG image itself is then included as string in the JSON message to the server using base64 encoding. This again adds an overhead of $33\%$, but ensures maintainability and extensibility as everything is still JSON. Furthermore would adding a second binary channel make the server more complex.

For applications using hardware accelerators we noticed that although the acceleration device itself is working to capacity, the host idles most of the time. The for the evaluation used GPU accelerated plasma simulation PIConGPU, e.g., uses only one CPU per GPU for busy waiting for events of other ranks. As most compute nodes have more CPU cores than dedicated GPUs, this leads to a poor utilization of CPU resources.

To exploit this, ISAAC tries to do as much CPU tasks in parallel as possible. Unfortunately most CPU tasks of ISAAC involve MPI. Even the concurrent rendering needs possible changed scene settings propagated over MPI beforehand. The MPI standard does in fact define ways to use it on the same rank in different threads, but even if vendor support is given, the performance is worse than without threading. We assume that this is because of synchronization overhead between the concurrent MPI calls. Because of this, only the compressing and sending of the image is done in background on the root rank while the application continues.

\begin{figure}
	\centerline{
		\resizebox{0.8\textwidth}{!}{\includegraphics{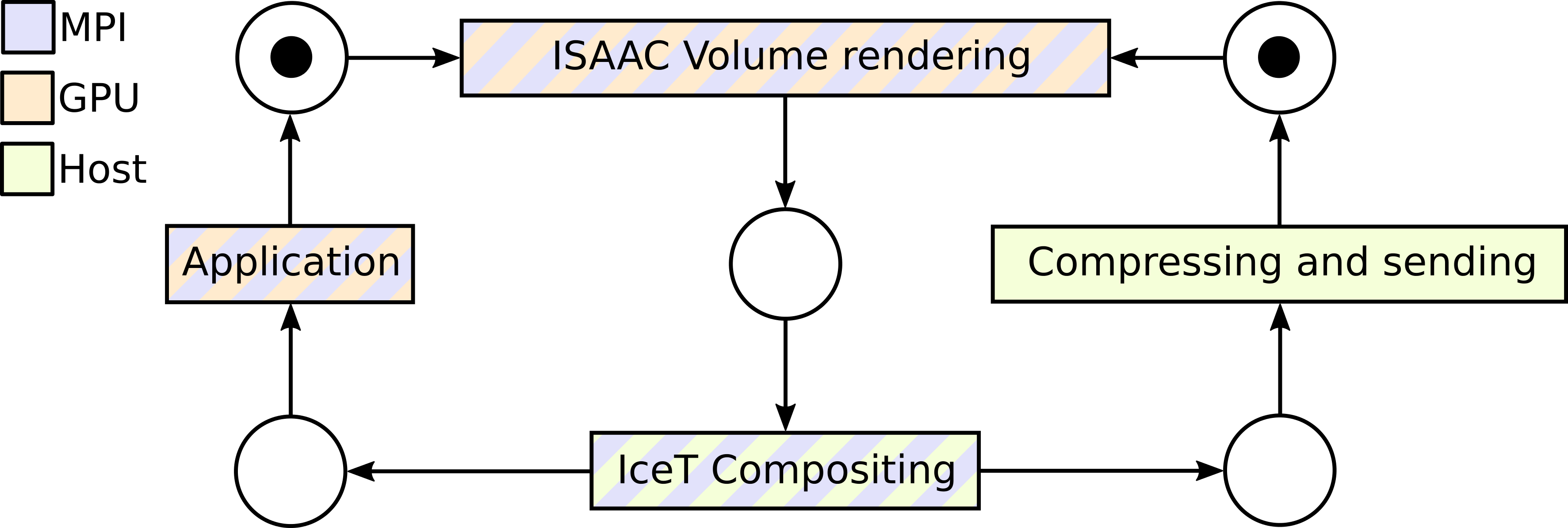}}
	}
	\caption{As ISAAC and the HPC application work on the same hardware accelerator and both use MPI, they cannot be executed in parallel. Only compressing and sending the final image is invariant of the application and MPI and can run in background while the application continues. Only if both concurrent tasks finish, the next frame can be rendered.}
	\label{fig:petri}
\end{figure}

The Petri net in figure \ref{fig:petri} depicts this. The scene setting propagation, the volume rendering and the IceT compositing need to pause the application, but the compression and sending can be done in parallel to the application. Only if both concurrent tasks have finished, the transition to render the next frame is activated again. If an application step is much faster than compressing and sending the image, a solution is to call ISAAC only every $n$th frame which may result in a lower framerate but increases the utilization of the whole system.

%% file: content/evaluation.tex
\section{Evaluation}

To demonstrate that ISAAC works well together with existing applications running on modern HPC systems (especially those equipped with hardware accelerators), it was used to visualize and steer the plasma simulation PIConGPU running on Nvidia GPUs.
The simulation code defines a plugin interface which enables user defined code to extend the simulation without the need of changing the core code. Just three new classes needed to be written: Two for the two possible field types of PIConGPU and one plugin class. Besides rendering a live visualization, also some meta data, like the particle count, are sent to the clients and the plugin listens to those enabling them to pause or even exit it.

\subsection{Run Time on a Petascale System}

We ran PIConGPU together with ISAAC on the supercomputer Piz Daint of the Swiss National Supercomputing Centre on up to 4096 Nvidia Tesla K20X GPUs \cite{MHW+16}. The cluster itself consists of 5272 GPUs, but because of availability and single user scheduling policies only 4096 were accessible for these tests. Each GPU has a single precision theoretical peak performance of $3.95$ TFLOP/s. All used 4096 GPUs together reach a peak performance of $\sim 16.2$ PFLOPS/s. PIConGPU is memory bound, but still capable to use over $12 \%$ of the single precision peak performance on the Kepler architecture~\cite{zenker2016performance}. On the investigated sub set of Piz Daint this means $\sim 1.9$ PFLOP/s are actually executed.

A Kelvin-Helmholtz instability is simulated. The simulation was parameterized with the Boris pusher, Esirkepov current solver, Yee field solver, trilinear-interpolation in field gathering, three spatial dimensions (3D3V), 128 cells in each dimension, electron and ion species with each sixteen particles per cell, and quadratic-spline interpolation (TSC)~\cite{hockney1988computer}. PIConGPU was compiled with nvcc 7.0 using the compiler flags \texttt{--use\_fast\_math --ftz=false -Xcompiler=-pthread -m64 -O3 -g}. With increasing the number of GPUs the local domain stays the same, but the global domain grows (weak scaling). The resolution of the ISAAC rendering was $1920 \times 1080$ (Full HD), interpolation was activated, two sources (electric field and current density) were rendered with complex functor chains including a square root operation, and 26 different view angles chosen such that well and poor cacheable memory accesses happened.

\begin{figure}[t]
	\centerline{
	\includegraphics{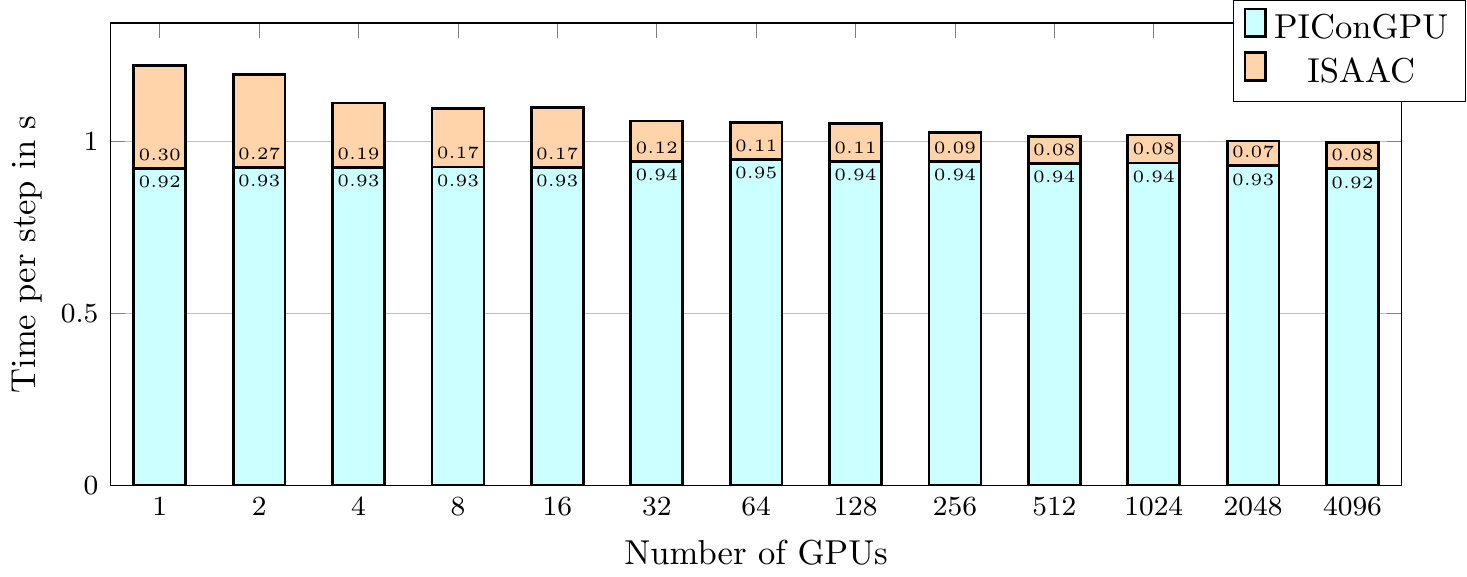}
	}
	\caption{PIConGPU with ISAAC running on up to 4096 Nvidia Tesla K20X GPUs ($\sim 16.2$ PFLOPS/s theoretical peak performance, single precision) on Piz Daint simulating a Kelvin-Helmholtz instability. Every GPU node simulates a volume of $128^3$. Adding more GPUs increases the global, simulated volume. PIConGPU shows perfect weak scaling~\cite{bussmann2013radiative}, while ISAAC gets faster as the resolution of the image per GPU shrinks and thus the number of needed ray casting rays.}
	\label{fig:piz_daint}
\end{figure}

Figure \ref{fig:piz_daint} shows the results of the runs. PIConGPU showed perfect weak scaling and always needed around $0.93$ seconds per time step independent from the number of GPUs. ISAAC on the other hand decreased the run time depending on the count of GPUs. The reason is that with a growing number of GPUs, the local image size decreases. There is not even a big difference between $512$ and $4096$ GPUs anymore: The rendering time itself reaches its minimum and the execution time is dominated by the render preparation and the IceT compositing which cannot be decreased easily anymore.

\begin{table}[t]
	\centering
	\begin{tabular}{|c|c|c|c|c|c|}
	\hline Number of GPUs & 1 & 8 & 64 & 512 & 4096 \\ 
	\hline Average runtime & 300 ms & 171 ms & 107 ms & 77 ms & 76 ms \\ 
	\hline Minimum runtime & 212 ms & 88 ms & 61 ms & 51 ms & 53 ms \\ 
	\hline Maximum runtime & 369 ms & 229 ms & 131 ms & 98 ms & 102 ms \\ 
	\hline 
	\end{tabular}
	\vspace{3mm}
	\caption{Average, minimum, and maximum runtime for 1, 8, 64, 512, and 4096 GPUs. Due to inefficient caching for some view angles the worst value may be up to three times slower than the best.}
	\label{tab:piz_daint_variance}
\end{table}

To show the effect of the 26 different camera angles, table \ref{tab:piz_daint_variance} shows the average, best, and worst runtime for 1, 8, 64, 512, and 4096 GPUs. The worst value may be up to three times slower than the best for some cases. This is because of the already mentioned inefficient caching as the ray casting algorithm may move perpendicular to the cache lines using only a few or even only one element of them.

\begin{figure}[t]
	\centerline{
	\includegraphics{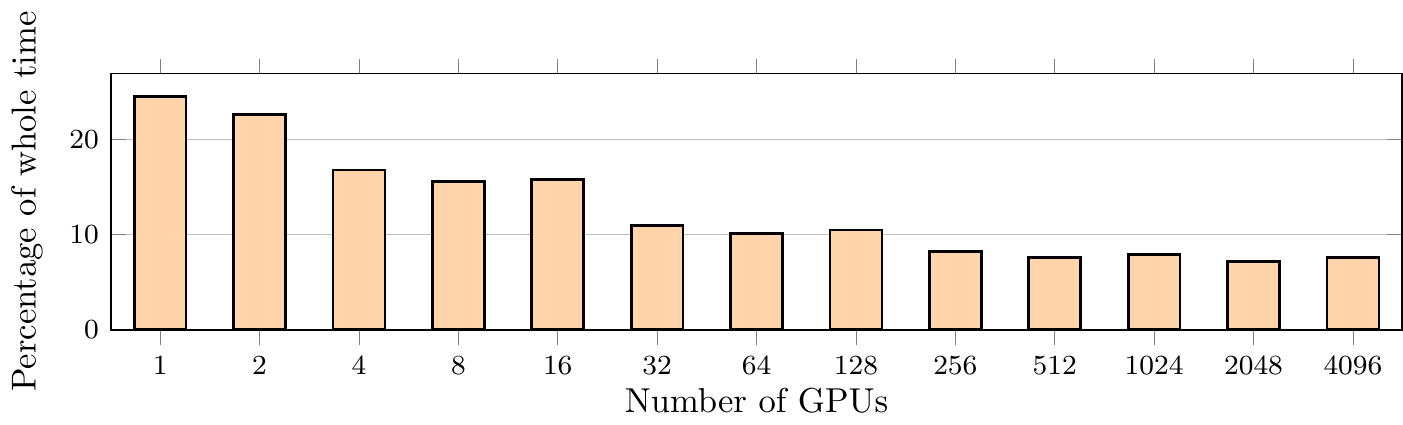}
	}
	\caption{Percentage of ISAAC run time compared to the total run time of the simulation run. With decreasing number of GPUs, ISAAC settles down at $\sim 8 \%$.}
	\label{fig:piz_daint_relative}
\end{figure}

Figure \ref{fig:piz_daint_relative} shows the run time of ISAAC compared to the total run time of the simulation in dependence of number of GPUs used. It shows that ISAAC is capable of visualizing a petascale application running on modern hardware accelerators without interrupting it for a too long time: From 256 GPUs on, ISAAC needs less than $10 \%$ of the total run time and finally settles down at $\sim 8 \%$.

The parameters of ISAAC were thereby set up quite conservative. Enabling interpolation creates better visualization if iso surface rendering is used, but may be deactivated for a visualization as glowing gas for performance reasons. The chosen resolution shows that Full HD (resolution of $1920 \times 1080$) renderings are possible, but most of the time smaller resolutions will fit the needs anyway and will be chosen as other client elements also need to fit on the screen. Accordingly the run time can still be decreased easily if needed.

\subsection{PIConGPU Renderings}

This section will give some examples of PIConGPU renderings and why they are useful for the users.

\begin{figure}[t]
	\centerline{
		\resizebox{1.0\textwidth}{!}{
			\includegraphics{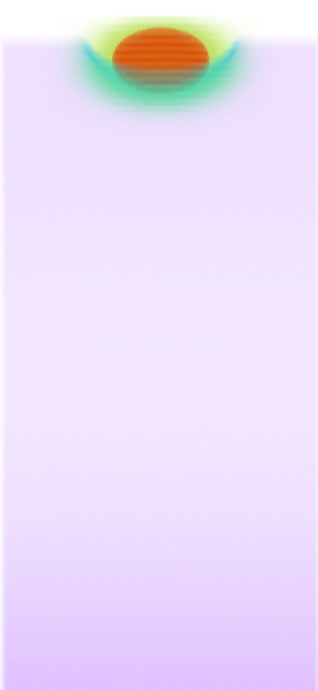}
			\includegraphics{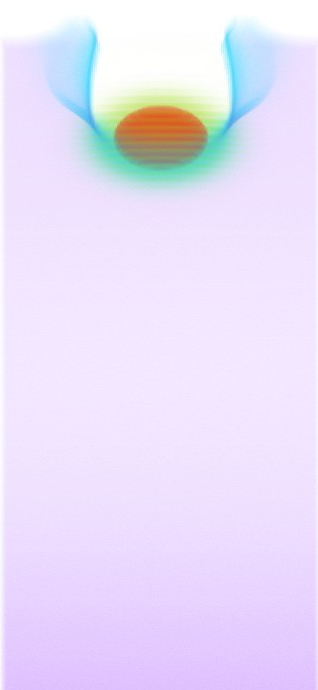}
			\includegraphics{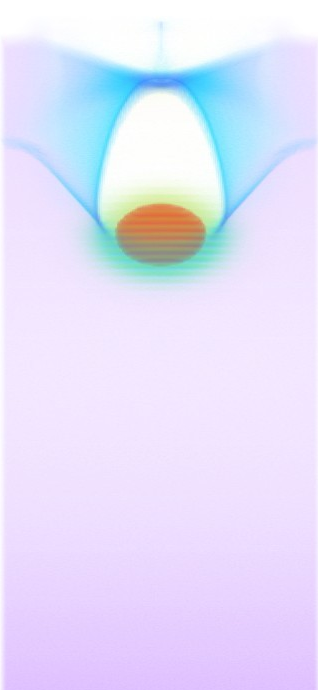}
			\includegraphics{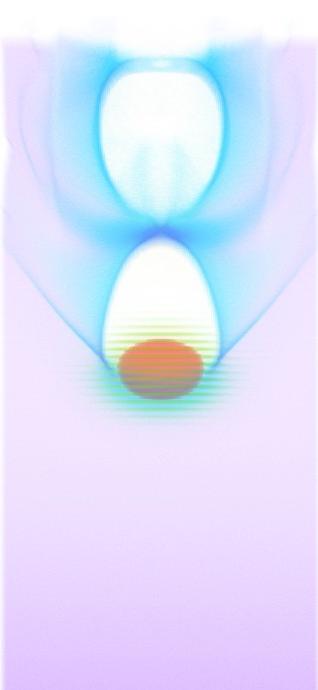}
			\includegraphics{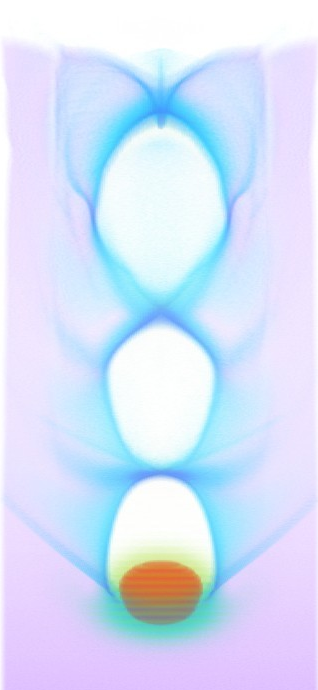}
			\includegraphics{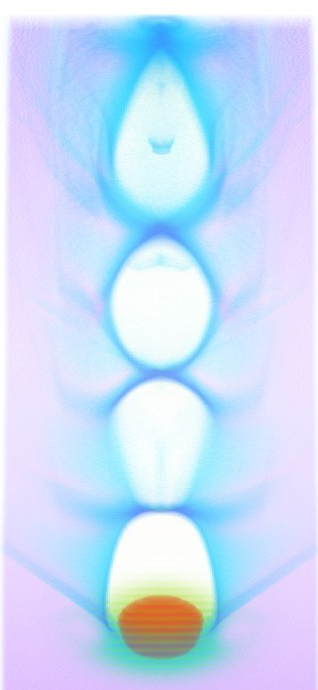}
			\includegraphics{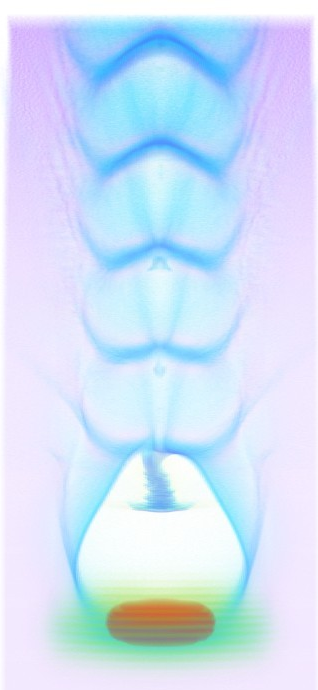}
			\includegraphics{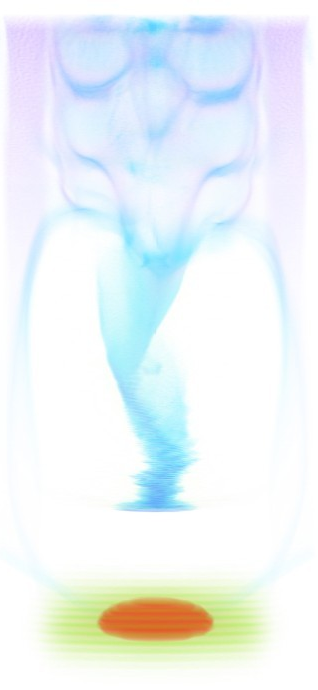}
		}
	}
	\caption{Time series of a laser wakefield accelerator}
	\label{fig:rendering:lwfa_seq}
\end{figure}

Figure \ref{fig:rendering:lwfa_seq} shows a time series of a so called laser wake field accelerator. A laser pulse (red and green) is ionizing and penetrating a gas. The electric-magnetic field of the laser pulse pushes the electrons (purple) off their ions (not shown). The moving electrons create a current which is shown in blue. As ions and electrons are seperated now, a plasma is created. Behind the laser pulse a bubble without electrons is created which is thereby positively charged as only inert ions remain. An electron can now be injected in this region as it can be seen in the both last images of the series, in which electrons from the back border enter the zone which is called self injection.

A physicist can see quite soon whether the simulation parameter show the expected phenomena, how long it takes until a self injection is happening, how the shape of the wake looks like, and how it changes over time. 

\begin{figure}[t]
	\centering
	\begin{minipage}{0.49\textwidth}
		\resizebox{\textwidth}{!}{\includegraphics{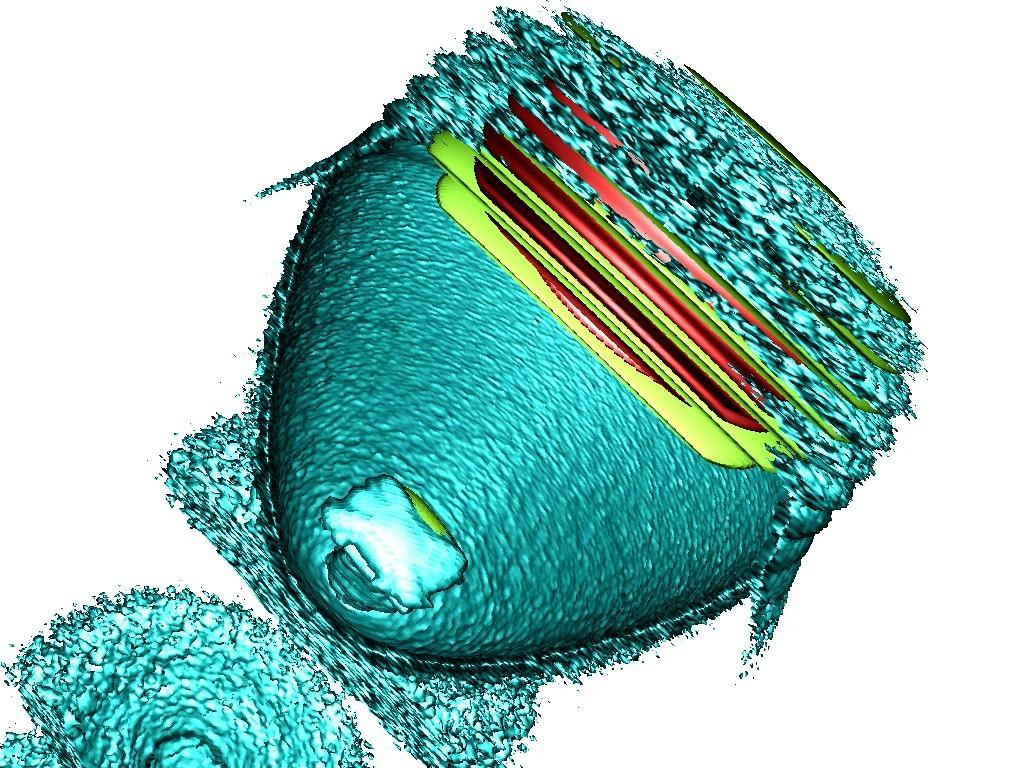}}
		\centerline{(a) Laser wakefield Accelerator}
	\end{minipage}
	\begin{minipage}{0.49\textwidth}
		\resizebox{\textwidth}{!}{\includegraphics{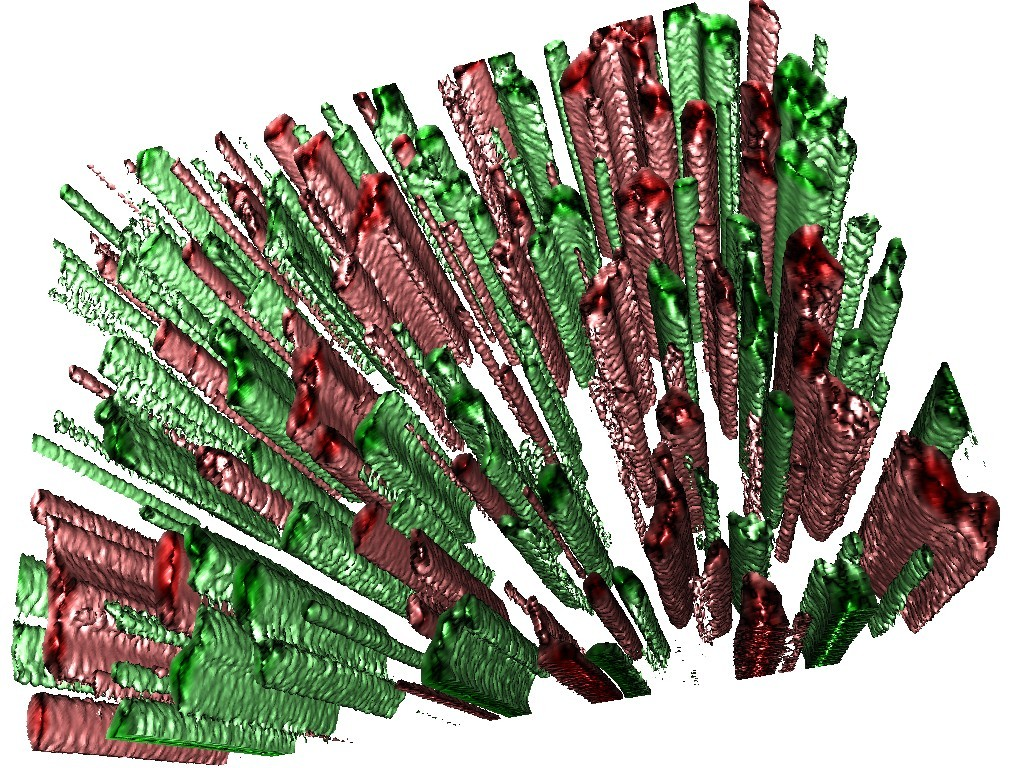}}
		\centerline{(b) Two stream instability}
	\end{minipage}
	\vspace{3mm}
	\caption{Laser wakefield accelerator and two stream instability with iso surface rendering}
	\label{fig:rendering:lwfa_weibel}
\end{figure}

Beside the visualization as glowing gas as seen in figure \ref{fig:rendering:lwfa_seq}, ISAAC does also support visualization with iso surfaces as seen in figure \ref{fig:rendering:lwfa_weibel}. In (a) the bubble behind the laser pulse can be seen directly as 3D object and even the self injection and its shape is easily recognizable. The picture (b) shows a so called two stream instability, in which plasma flows in opposite directions creating long standing filaments and magnetic fields. While the simulation is running, the user can see the tubes appear, disappear and change their shape. The simulation can be paused at any time, the visualization parameters changed, a record of data started in the simulation or more (meta) data requested to be sent to the client. Going back in time would be an option, too, if the simulation supports this, e.g. with loading a slightly older checkpoint.

\begin{figure}[t]
	\centerline{
		\resizebox{1.0\textwidth}{!}{\includegraphics{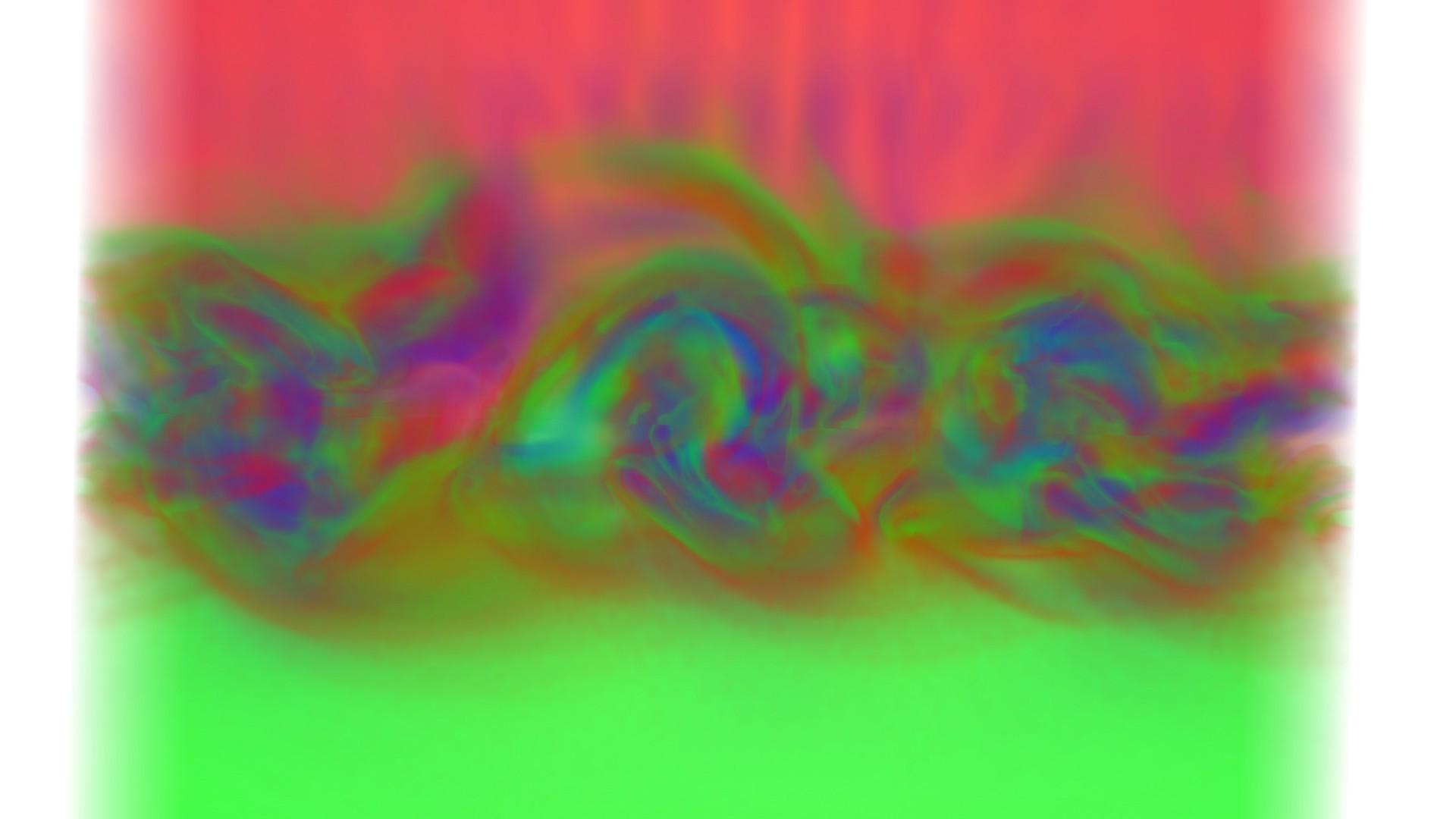}}
	}
	\caption{Kelvin-Helmholtz instability}
	\label{fig:rendering:khi}
\end{figure}

Using different colors and color maps may also help understanding the behavior of the simulation. Figure \ref{fig:rendering:khi} shows a Kelvin-Helmholtz instability, where plasma flows in two streams side by side with different speeds, whereas vortexes are created. The green and red color identify both electrons, but with every color showing a different stream. The vortex effect can easily be spotted and observed over time. Furthermore, the electric field created by the moving electrons is shown in blue.

%% file: content/conclusion.tex
\section{Conclusion}

In this paper we presented ISAAC, an open-source library for the in situ visualization and steering of applications running on HPC computers using multi-core CPUs or hardware accelerators like Nvidia GPUs. It is designed to work on the original application data structures and types without the need of deep copies or data format conversions. It provides volume rendering visualization as well as iso surface rendering and supports clipping planes, free transfer functions, and data interpolation for the fine tuned control of the output image.

Beside live rendering we showed that it is possible with ISAAC to ship arbitrary meta data packed in the open JSON format and to send feedback back to the application. A central server manages connections from supercomputer applications and clients and forwards messages between them as well as creates streams out of the raw pictures reducing the needed bandwidth. The central server enables to explore these applications even from outside the site network.

For demonstrating the capabilities of ISAAC on a real world application, we added it to the world fastest particle-in-cell code PIConGPU~\cite{bussmann2013radiative} and ran it on the supercomputer Piz Daint. We could show that ISAAC is capable of visualization such a petascale simulation with only using $8 \%$ of the compute time for high quality Full HD images. We provided examples how ISAAC's live visualizations helps scientists on three different plasma effects simulateable with PIConGPU.

It is left for future work to test ISAAC on other large-scale high performance compute systems and hardware platforms. We expect the performance of ISAAC on a system to mainly depend on the scalabilty and performance of the Alpaka and IceT libraries used.